\documentclass[journal]{IEEEtran}

\usepackage[utf8]{inputenc}
\usepackage[T1]{fontenc}
\usepackage{amsmath,amssymb,amsfonts}
\usepackage[ruled, linesnumbered, noend, french, onelanguage]{algorithm2e}
\usepackage{graphicx}
\usepackage{subcaption}
\usepackage{url}
\usepackage{longtable}
\usepackage{pdfpages}
\usepackage[shortlabels]{enumitem}
\newcommand{\bC}{\mathbb{C}}

\newcommand{\bE}{\mathbb{E}}

\newcommand{\bN}{\mathbb{N}}

\newcommand{\bR}{\mathbb{R}}

\newcommand{\cE}{\mathcal{E}}

\newcommand{\cL}{\mathcal{L}}

\newcommand{\cP}{\mathcal{P}}

\newcommand{\dx}{\text{d}}

\newcommand{\latent}[1]{h_{#1}}

\newcommand{\observation}[1]{x_{#1}}
\newcommand{\estimation}[1]{\hat{x}_{#1}}

\newcommand{\Fs}{F_s}

\newcommand{\Hs}{G_s}

\newcommand{\us}{u_s}
\newcommand{\vs}{v_s}

\newcommand{\thh}{a}
\newcommand{\thx}{b}
\newcommand{\txh}{c}
\newcommand{\vh}{\alpha}
\newcommand{\vx}{\beta}
\newcommand{\vi}{\eta}

\newcommand{\gaussienne}{\mathcal{N}}

\newcommand{\dirac}[2]{\delta_{#1}(#2)}
\newcommand{\var}[1]{\text{ Var}(#1)}

\newcommand{\cov}[2]{\text{ Cov}(#1,#2)}

\newcommand{\parameters}{\theta}
\newcommand{\p}{p_\parameters}
\newcommand{\f}{f_\parameters}

\newcommand{\dimh}{n}
\newcommand{\hankel}{\mathcal{R}_\infty}
\newtheorem{proposition}{Proposition}

\newtheorem{lemma}{Lemma}
\newtheorem{theorem}{Theorem}

\newsavebox{\mybox}

\ifCLASSINFOpdf
\else
   \usepackage[dvips]{graphicx}
\fi
\usepackage{url}

\usepackage{graphicx}

\usepackage{tikz-cd}
\usepackage{tikz}

\usetikzlibrary{positioning,shapes,shadows,arrows, automata}
\usetikzlibrary{decorations.pathmorphing}
\usetikzlibrary{decorations.markings}
\definecolor{myred}{RGB}{200,10,10}

\tikzstyle{symbol} = [shape=rectangle,rounded corners = 5pt,
draw, align=center, minimum width=1.8em, minimum height = 1.5em,
top color=white]
\tikzstyle{passive} = [shape=circle,
draw, align=center, minimum width=2em,
top color=white]
\tikzstyle{active} = [shape=circle,
draw, align=center, minimum width=2em,
top color=white, bottom color=red!20]
\tikzstyle{machine} = [shape=rectangle,
draw, align=center, minimum width=2em,
top color=white]
\tikzstyle{log} = [shape=ellipse,
draw, align=center, minimum width=1em,
top color=white, bottom color=yellow!20]
\tikzstyle{tdep} = [color = myred, ultra thick,->]
\tikzstyle{sdep} = [color = mygreen, ultra thick,->]
\tikzstyle{alarm} = [shape=rounded rectangle,
draw, align=center, minimum width=2em,
top color=white]

\newtheorem {remark} {Remark}

\begin{document}

\title{Expressivity of Hidden Markov Chains vs. Recurrent Neural Networks from a system theoretic viewpoint}

\author{Fran\c{c}ois Desbouvries, \IEEEmembership{Senior Member, IEEE}, Yohan Petetin, \IEEEmembership{Member, IEEE}, and Achille Sala\"{u}n
\thanks{Fran\c{c}ois Desbouvries and Yohan Petetin are with Samovar, Telecom SudParis, Institut Polytechnique de Paris, Evry, France (e-mail: francois.desbouvries,yohan.petetin@telecom-sudparis.eu).
Achille Sala\"{u}n is with Institute of Biomedical Engineering, Department of Engineering Science, University of Oxford, Oxford, UK. The work was performed when he was with Samovar, Telecom SudParis, Institut Polytechnique de Paris, Evry, France, and Nokia Bell Labs, Nozay, France (e-mail: achille.salaun@eng.ox.ac.uk).}}

\markboth{ Vol. xxx, No. xxx, xxx 2021}
{S
hell \MakeLowercase{\textit{et al.}}: Bare Demo of IEEEtran.cls for IEEE Journals}
\maketitle

\begin{abstract}
Hidden Markov Chains (HMC) and 
Recurrent Neural Networks (RNN)
are two well known tools for predicting time series. 
Even though these solutions were developed independently in distinct communities, they share some similarities 
when considered as probabilistic structures. 
So in this paper we first consider HMC and RNN as generative models, and we embed both structures in a common generative unified model (GUM).
We next address a comparative study of the expressivity of these models. 
To that end we assume that the models are furthermore linear and Gaussian.
The probability distributions produced by these models are characterized by structured covariance series, 
and as a consequence expressivity reduces to comparing sets of structured covariance series, 
which enables us to call for stochastic realization theory (SRT).
We finally provide conditions under which a given covariance series can be realized by a GUM, an HMC or an RNN.
\end{abstract}


\IEEEpeerreviewmaketitle

\section{Introduction}
Let $x_{0:t}=(x_{0},\cdots,x_t)$ be a sequence of random variables (r.v.).
We focus on the general problem of predicting
a future observation $x_{t+1}$ from a realisation of $x_{0:t}=(x_{0},\cdots,x_t)$. 
This problem has many applications such as speech recognition,
finance or geology \cite{shumstof2000}\cite{douc2014nonlinear}
and can be addressed through Bayesian estimation in two ways.
The first way consists in estimating a generative model $\p(x_{0:t})$, for all
$t \in \bN$, and next computing the posterior distribution
$\p(x_{t+1}|x_{0:t})$. The second approach aims
at building directly a function $\f$ such that
$\f(x_{0:t})$ is close to $x_{t+1}$ is a given sense.
The objective of this paper is to propose a comparison between
two key tools associated with each approach, hidden Markov Chains (HMC) on the one hand,
and recurrent neural architectures (RNN)
on the other hand.
Our study is {\sl not} of an experimental nature
(see e.g. \cite{Deshmukh_2020}\cite{hmm-rnn-traffic} for such comparisons),
but rather aims at quantifying the modeling power of each model.
Before further comparing these two models,
let us briefly review the rationale of the
two approaches by recalling
the prediction problem in the static case.

\subsection{Bayesian problem}
Let us consider a sample $(x,y)$ from a joint probability density function (pdf) $p(x,y)$.
The objective is to predict $y$ from $x$,
so we look for an estimator $\hat{y}=f(x)$ such that 
$\hat{y}$ is  "close" to $y$.
In a Bayesian context, building estimator $f(.)$ is induced by the choice of a loss function $L(.,.)$, which depends on the problem at hand, and quantifies the error
between the prediction $f(x)=\hat{y}$ and the true variable $y$.
Building the associate estimator amounts to minimizing the Bayesian risk
\begin{equation}
\label{bayesian-risk}
R(f) = \bE\left[L(f(x),y)\right],
\end{equation}
i.e. build $f^\star (x)=\hat{y}$ in which
$f^\star = \underset{f}{\text{argmin }} R(f)$.
One can show that 
$f^\star$ depends on the \textit{posterior} density
$p(y|x)=\frac{p(x,y)}{p(x)}$ 
(also called \textit{predictive distribution} in the context of prediction). For instance,
if the loss is quadratic,
the Bayesian estimator is well known to be the conditional expectation,
$f^\star(x) = \hat{y}=\bE[y|x]$.
However, $p(x,y)$ is not known in practice.
To cope with this problem one can estimate the Bayesian risk %\eqref{eq:minL} 
by two different ways:
by introducing a parameterized distribution 
$p_\parameters(x,y)$, 
or by estimating integral \eqref{bayesian-risk}
from Monte Carlo samples. 

\subsubsection{Parameterizing the joint distribution $p(x,y)$}
The first approach consists in proposing a model 
of the unknown pdf $p$.
We thus restrict ourselves to a parameterized set of pdfs $\left(p_\theta\right)_{\parameters\in\Theta}$,
in which $\parameters$ can be multidimensional.
If we have a set of labelled independent samples
\begin{align}
\cE=\left\{(x_i,y_i)\overset{\text{i.i.d.}}{\sim}p(x,y)\right\}_{1 \leq i \leq n}
\label{eq:data},
\end{align}
the relevance of $p_\parameters$ can be quantified via the likelihood function
\cite{borovkov1987statistique,wasserman2013all}
\begin{align}
    \cL(.;\cE):\parameters\mapsto\prod_{i=1}^n p_{\parameters}(x_i,y_i),
    \label{eq:likelihood}
\end{align}
so approximating $p$ amounts to computing a parameter $\parameters$ which maximizes the likelihood.
Note however that the choice of the parametric family is critical:
$p_\parameters$ should model the data at hand, 
and in the same time
function
\eqref{eq:likelihood}
should be computed and optimized efficiently.
In general, maximizing the likelihood can only be done approximately.
For instance in the case of latent variables models
(i.e. the model is defined through the introduction of an unobserved random variable $h$ such that $p_\parameters(x,y)=\int p_\parameters(h,x,y){\rm d}h$), 
maximizing the likelihood requires approximating schemes 
such as 
the Expectation Maximization
(EM) algorithm
\cite{dempster1977em}.

\subsubsection{Parameterizing the estimator $f(x)$}
The second approach does not make any assumption on 
$p(x,y)$, 
but rather estimates \eqref{bayesian-risk}
from the same dataset \eqref{eq:data}
(see e.g. \cite{bishop2006pattern,hastie2009elements}). 
The problem of building an estimator becomes that of minimizing the empirical risk
\begin{align}
    f^\star_n=\underset{f}{\text{argmin }}\frac{1}{n}\sum_{i=1}^n L(f(x_i),y_i).
    \label{eq:minf}
\end{align}
However, since the dataset is finite,
in the absence of further constraints,
any function interpolating the points $(x_i,y_i)$
satisfies the optimisation problem \eqref{eq:minf}.
In such a case, the model overfits 
and proves unable to generalize to new observations.
This problem is often overcome by chosing a family of functions $(f_\parameters)_{\parameters\in\Theta}$,
and finally \eqref{eq:minf} turns into the parameter estimation problem:
\begin{eqnarray}
\parameters^\star_n
&=&
\underset{\parameters}{\text{argmin }}R_n(f_\parameters),
\label{eq:minftheta}
\\
R_n(f_\parameters)
&=&
\frac{1}{n}\sum_{i=1}^n L(f_\parameters(x_i),y_i)
\end{eqnarray}
which eventually produces the estimator
$\hat{y}=f_{\parameters^\star_n}(x)$ (notation $\parameters^\star_n$
underlines the fact that the estimator depends on the training set 
$\cE$, which is of dimension $n$).

As above, the choice of the family $(f_\parameters)_{\parameters\in\Theta}$
should be balanced:
a poor set of functions will lead to unrealistic predictions,
while a rich set of functions can lead to overfitting.
Moreover $(f_\parameters)_{\parameters\in\Theta}$
should lead to tractable learning, 
i.e. it should be possible to solve \eqref{eq:minftheta}
efficiently.
Classical solutions include
the functions belonging to 
a reproducing kernel Hilbert space (RKHS) 
\cite{manton2014primer}
\cite{paulsen2016introduction}
and the functions defined by a neural network (NN)
\cite{jain1996artificial}
\cite{lecun2015deep}.
Optimizing \eqref{eq:minftheta} for these families of functions leads to well known algorithms
such as (linear or kernel based) least squares
\cite{bishop2006pattern},
Support Vector Machines (SVM)
\cite{burges1998tutorial}
\cite{hu2003robust}
\cite{vapnik2013nature},
or deep learning algorithms 
\cite{bishop2006pattern} \cite{goodfellow2016deep}
for regression or classification.

\subsubsection{Discussion}
As we have just seen,
for minimizing \eqref{eq:minftheta} 
it is not necessary to mimic 
the distribution $p(x,y)$ by $\p(x,y)$.
In addition, under some assumptions about the family $(f_\theta)_{\theta\in\Theta}$ 
it is possible to derive concentration inequalities such as
\begin{equation*}
%\label{eq:inequality-1}
\mathbb{P} \left(|R_n(f_{\theta^\star_n})-R(f_{\theta^\star_n})|>\epsilon \right) \leq \delta_{\epsilon,n} \text{,}
\end{equation*}
where $\delta_{\epsilon,n} \rightarrow 0$ when
$n \rightarrow \infty$ \cite{Vapnik1998}; 
in other words, such a bound ensures that 
$f_{\theta^\star_n}$ generalizes well and also provides
a rate of convergence.
However, note that in some contexts,
and in particular for times series analysis, 
we may be interested in predicting $\phi(x)$ from $y$  for a large class of functions $\phi$; 
once the joint distribution has been estimated, 
it is possible to comply 
with such a constraint without running a new estimation algorithm for each function $\phi$. In addition, the knowledge of the posterior distribution enables to quantify (even approximately) the uncertainty of the prediction. 

\subsection{Goal of this paper}
The two previous
approaches can be adapted to the sequential constraints induced by time series analysis.
Let $t$ be the current time parameter.
In order to represent the joint distribution of $x_{0:t}$, 
we need to choose a parametric generative model 
$\p(x_{0:t})$ such that $\theta$ does not depend on $t \in \bN$ (otherwise, the model cannot be used with new observations). $\p$ should model the time series $x_{0:t}$ in a realistic way,
and so take into account the dependencies between the observations; in the same time, we should be able to compute an estimator of $\theta$ and to approximate the posterior distribution $\p(x_{t+1}|x_{0:t})$ for any realization $x_{0:t}$.

A popular model satisfying these requirements is the HMC, particularly developed in the signal processing community,
see e.g.
\cite{rabiner1986introduction}
\cite{rabiner1989tutorial}.
In the same way, it is possible to parameterize a function $f_{\theta}(x_{0:t})$ in a such way that $\theta$ does 
not depend on $t$. This is the principle of RNN particularly developed in the
machine learning community \cite{connor1994recurrent}.
Even if such models were basically proposed for point estimation, they can be easily used for building generative models. 
So from now on, in order to compare the two approaches in a common framework, we will consider that
we have at our disposal two generative models $\p(x_{0:t})$ for all $t$, the HMC and the RNN. 
Starting from the observation that both models actually rely on a set of 
latent variables and that they share some common features
in the construction of these variables, our objective in this paper is to quantify thoroughly (under some assumptions)
how the structural differences of these models impact on their expressivity. 

The rest of this paper is organized as follows. 
In Section \ref{subsec:motiv_pred}
we start by formalizing both models under a common framework.
Next in section \ref{structured-mapping}, 
we see that comparing both models 
under the linear and Gaussian stationary assumptions
reduces to comparing the covariance series of the stochastic process $x_{0:t}$,
and consequently the study calls on
Stochastic Realization Theory (SRT)
(a branch of systems theory).
Section \ref{app:trs} provides a brief summary of SRT.
Finally the expressivity of HMC and RNN is compared in section \ref{expressivity-GUM-HMM-RNN}.

\section{Latent data models for time series analysis}
\label{subsec:motiv_pred}
%\label{sec:generative}

In this section we introduce our two generative models based
on a sequence of latent r.v. $h_{0:t}$, and we next
embed them as two particular instances of a 
more general model.

\subsection{Markovian models}\label{sec:markov}

When dealing with a time series $x_{0:t}$ one major 
issue consists in modeling the dependency between
the observations. 
For example, a simple Markov Chain (MC)  
\begin{align}
\label{eq:facto}
p_\parameters(\observation{0:t})
=p_\parameters(\observation{0})\prod_{s=0}^{t-1} p_\parameters(\observation{s+1}|\observation{s})
\end{align}
is often unlikely to represent the distribution of $x_{0:t}$ in a realistic way ,
since $x_t'$, $t'<t-1$, becomes independent of $x_t$ 
when $x_{t-1}$ is observed.
One way of enhancing expressivity
is to introduce a latent process $\latent{0:t}$, 
where each 
$h_s$ can be discrete or continuous.
The model is now described by the full joint density 
$p(\latent{0:t},\observation{0:t})$,
from which $p(\observation{0:t})$
is obtained by marginalizing out the latent variables. %$\latent{0:t}$
This marginalization definitely makes
pdf $p(\observation{0:t})$ more complex and so increases the modeling power.
A constraint is that the introduction of a latent process should preserve some computational properties in order to be used in practice.
In this sense, the HMC generalizes \eqref{eq:facto}
by adding a latent process in a rather parsimonious way, since its joint pdf reads:
\begin{align}
\label{eq:joint-hmc}
p_\parameters(\latent{0:t}, \observation{0:t}) 
\overset{\text{HMC}}{=}  p_\parameters(\latent{0})\prod_{s=1}^{t} p_\parameters(\latent{s}|\latent{s-1})
\prod_{s=0}^{t}p_\parameters(\observation{s}|\latent{s}).
\end{align}
So an HMC benefits of three (conditional) independence properties:
the latent process is an MC;
given the latent variables 
$\latent{0:t}$,
observations 
$\observation{0:t}$
are independent;
and given all latent variables 
$\latent{0:t}$,
an observation only depend on the latent variable at the same time, $p_\parameters(\observation{s}|\latent{0:t})=p_\parameters(\observation{s}|\latent{s})$ for all $s$, $0 \leq s \leq t$.
The hidden process $h_{0:t}$
can have a physical 
meaning
(in which case estimating 
$h_{0:t}$ from
$x_{0:t}$ is relevant).
If not,
the role of the latent variables $h_{0:t}$
is just to make the observed process
$x_{0:t}$ more complex,
and the HMC can be seen as a generative model.
Associated inference algorithms for approximating the Maximum Likelihood estimator and  posterior distributions have been extensively studied for these models \cite{cappe2005inference} \cite{douc2014nonlinear} and are recalled 
in Appendix \ref{Computations-HMM}.

\subsection{RNN architectures}
\label{architectures-neuronales}
RNN are an adaptation of neural architectures
to times series. So let us start by briefly recalling
the rationale of neural networks.

Neural network architectures are versatile classes of functions
\cite{goodfellow2016deep},
which have found many applications for classification or prediction, including language  \cite{mikolov2011extensions} or image  \cite{krizhevsky2012imagenet}
processing.
A neural network (NN) is a succession of parameterized functions called neurons. A neuron typically computes 
$\boldsymbol{x}\mapsto\sigma(\boldsymbol{w}\boldsymbol{x}+b),
$
where 
$\boldsymbol{w} \boldsymbol{x}$ is the dot product of 
$\boldsymbol{w}$ 
(a vector of weights) and 
$\boldsymbol{x}$ (a vector of variables), 
$b$ is the bias,
and
$\sigma(.)$ is a so-called (nonlinear) activation function,
such as the sigmoid,
hyperbolic tangent or ReLu functions.
Neurons can be gathered into layers which themselves can be cascaded, yielding increasingly complex functions. 
Some universal approximation theorems have been proposed
\cite{cybenko1989approximation,hornik1991approximation,pinkus1999approximation,lu2017expressive};
for instance, given any (possibly multi-dimensionnal)
continuous function $f$,
there exists a single-layer NN
$f_\theta$ arbitrarily close to $f$, provided the activation function is not  
polynomial \cite{pinkus1999approximation}.
Similar results have been proposed for multiple layers NNs.
So any Lebesgue-integrable function $f:\bR^n\to\bR$
can be approximated by an NN with ReLu activation function and layers made of at least $n+4$ neurons,
provided the net is deep enough
\cite{lu2017expressive}.
The number of layers and of neurons per layer, 
as well as the activation functions,
are hyperparameters which characterize the NN architecture,
and the weights and biases are the model parameters
learnt from a training set.
However, the input of an NN as described above is of fixed size, which is not well suited to the modeling of time series in which observations accumulate - unless we use a sliding window, but in that case the prediction would not be based on the full set of observations.

In order to introduce dependencies between all the observations,
RNN introduce a latent variable $\latent{t}$
which is a function of all observations $\observation{0:t}$ and serves as a memory of the past.
After receiving the new information  $\observation{t}$, 
the new state is computed as 
\begin{equation}
\label{eq:latent_rnn}
\latent{t}=f_{\parameters}(\latent{t-1},\observation{t}) \text{,} \quad \latent{-1}=0
\end{equation}
where ${\parameters}$
is an NN layer.
In other words, $\latent{t}$ is a summary of all 
the past observations until time $t$.
Finally a prediction 
of $\observation{t+1}$ 
 is computed as
 $\estimation{t+1}=g_{\theta}(\latent{t})$
where $g_{\theta}$ is an NN architecture.
As we claimed before, RNN can be transformed into generative models by replacing function $g_{\parameters}$
by a parametric distribution $\p(x_{t+1}|h_{t})$.
In this case, we obtain 
a family of models defined by
\begin{align}
\label{eq:joint-rnn}
p_{\parameters}(\observation{0:t}) & \overset{\text{RNN}}{=} p_{\parameters}(\observation{0})
\prod_{s=0}^{t-1} p_{\parameters}(\observation{s+1}|\latent{s}),
\end{align}
where $\latent{s}$ depends on $\parameters$ 
from \eqref{eq:latent_rnn}.

By construction, the posterior distribution
$\p(x_{t+1}|x_{0:t})$ coincides with 
$\p(x_{t+1}|h_t)$ and is directly available.
The estimation of $\theta$ can be computed
by a gradient backpropagation algorithm
\cite{rumelhart1985learning}
\cite{robinson1987utility}
\cite{werbos1990backpropagation}
\cite{mozer1995focused}
which aims at maximizing $\log(\p(x_{0:t}))$ for a
given observation.
Due to the time component, 
there can be as many 
computed gradients as observations
for a given parameter.

However, in practice, the gradients computed for a given parameter
geometrically tend to
infinity or to zero
when we get back into the past.
These phenomena 
are called exploding
gradient and
vanishing gradient.
The exploding gradient phenomenon
is often due to the repeated multiplication of high weights,
a situation where learning the RNN becomes particularly unstable.
An efficient way to limit this behavior
is to bound the values taken by the gradient \cite{goodfellow2016deep,goldberg2017neural}.
One can also include a regularization term to the cost function
in order to penalize weights that are too large \cite{pascanu2013difficulty}.
By contrast, 
the vanishing gradient phenomenon results from the repeated  multiplication
or small size weights,
as well as the iterated use of activation functions which have derivatives bounded by 1 in magnitude
(\textit{e.g.} the sigmoid).
In that case, 
the oldest observations are not taken into account in the learning phase, 
so it is difficult to learn long term dependencies.
In order to mitigate the vanishing gradient phenomenon,
more sophisticated  architectures have been proposed,
such as the
Long Short Term Memories (LSTM) and the Gated Recurrent Units (GRU)
\cite{chung2014empirical}
(the only difference is that the corresponding parameterization of $f_{\parameters}$ 
becomes more complex).

\subsection{Generative unified model}
As we have just seen, HMC and RNN models
result from a different paradigm but both aim
at proposing a parameterized distribution $\p(x_{0:t})$ via
the introduction of latent variables $h_{0:t}$. 
In the RNN model, 
$h_{0:t}$ is deterministic given the observations $x_{0:t}$,
and $h_s$ summarizes all the observations up to time $s$ into a unique variable;
in the HMC model, $h_{0:t}$ is  stochastic given the observations, and indeed the Bayesian estimation of 
$h_{0:t}$ is of interest in cases where $h_{0:t}$
is a physical process of interest.

When we put aside the computational aspects, the natural question that arises is to compare the set of distributions
$\p(x_{0:t})$ induced by each model. 
Actually, both representations can be reconciled as particular instances of the following Generative Unified Model (GUM), 
\begin{equation}
\label{def-GUM}
\p(\latent{0:t},\observation{0:t})
\overset {\rm GUM} {=}
\p(\latent{0})
\prod_{s=1}^t \p(\latent{s}|\latent{s-1},\observation{s-1})
\prod_{s=0}^t \p(\observation{s}|\latent{s}).
\end{equation}
Indeed, the HMC model \eqref{eq:joint-hmc} is a GUM where
\begin{equation}
\label{eqcle-HMM}
\p(\latent{t}|\latent{t-1},\observation{t-1})
\stackrel{\rm HMC}{=}
\p(\latent{t}|\latent{t-1}) \text{,}
\end{equation}
while the RNN 
\eqref{eq:latent_rnn}-
\eqref{eq:joint-rnn} satisfies (up to the transformation $\latent{t}\leftarrow\latent{t-1}$)
\begin{align}
\label{eqcle-RNN}
\begin{aligned}
\p(\latent{t}|\latent{t-1},\observation{t-1})
&\stackrel{\rm RNN}{=}
\dirac{\f(\latent{t-1}, \observation{t-1})}{\latent{t}} \text{,} \\ \p(x_0|h_0) &\stackrel{\rm RNN}{=}\p(x_0) \text{,} \\ 
h_0&\stackrel{\rm RNN}{=}0 \text{,}
\end{aligned}
\end{align}
where $\delta$ denotes the Dirac mass. 
In the rest of this paper we will also consider deterministic GUM
(D-GUM), which are defined by the first equation of 
\eqref{eqcle-RNN}
only
(the interest of D-GUM over RNN will be clear in section \ref{expressivity-gum-hmm-rnn}).

Let us now discuss the three models,
beginning with their similarities.
First, in all three models the pair 
$(h_s,x_s)$ is an MC:
\begin{eqnarray}
p(h_s,x_s|h_{0:s-1},x_{0:s-1}) \!\!\!\!\!\!&=&\!\!\!\!\!\!
p(h_s,x_s|h_{s-1},x_{s-1})
\nonumber \\
\!\!\!\!\!\!&=&\!\!\!\!\!\!
p(h_s|h_{s-1},x_{s-1})
p(x_s|h_{s}),
\end{eqnarray}
which induces 
$p(x_s|h_{0:s},x_{0:s-1})=p(x_s|h_s)$.
In addition, in all three models the marginal process
$h_{0:t}$
is also an MC,
since
\begin{align*}
\p(\latent{s}|\latent{0:s-1}) &=\int \p(\latent{s}|\latent{s-1},\observation{s-1}) \p(\observation{s-1}|\latent{s-1}) {\rm d} \observation{s-1} \\ &=\p(\latent{s}|\latent{s-1}) \text{.}
\end{align*}

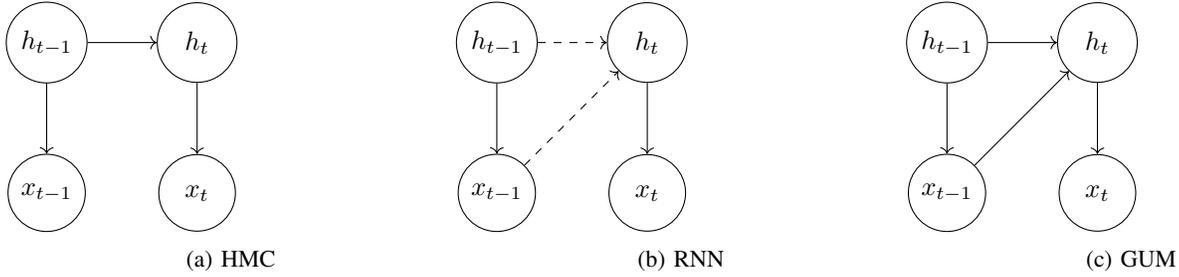
\begin{figure*}[h!]
    \begin{center}
	\begin{subfigure}[t]{.33\linewidth}
        \begin{tikzpicture}
            \node [passive, text width={width("$\observation{t-1}$")}] (h0) at (0,0) {$\latent{t-1}$}; 
            \node [passive, text width={width("$\observation{t-1}$")}] (h1) at (2,0) {$\latent{t}$}; 
            \node [passive, text width={width("$\observation{t-1}$")}] (x0) at (0,-2) {$\observation{t-1}$};
            \node [passive, text width={width("$\observation{t-1}$")}] (x1) at (2,-2) {$\observation{t}$};
            \draw[->] (h0) -- (h1); % , color=red
            \draw[->] (h0) -- (x0); % , color=red
            \draw[->] (h1) -- (x1);
        \end{tikzpicture}
		\caption{HMC}
	\end{subfigure}%\hfill
	\begin{subfigure}[t]{.33\linewidth}
        \begin{tikzpicture}
            \node [passive, text width={width("$\observation{t-1}$")}] (h0) at (0,0) {$\latent{t-1}$}; 
            \node [passive, text width={width("$\observation{t-1}$")}] (h1) at (2,0) {$\latent{t}$}; 
            \node [passive, text width={width("$\observation{t-1}$")}] (x0) at (0,-2) {$\observation{t-1}$};
            \node [passive, text width={width("$\observation{t-1}$")}] (x1) at (2,-2) {$\observation{t}$};
            \draw[->, dashed] (h0) -- (h1); % , color=red
            \draw[->] (h0) -- (x0); % , color=red
            \draw[->, dashed] (x0) -- (h1); % , dashed, color=red
            \draw[->] (h1) -- (x1);
        \end{tikzpicture}
		\caption{RNN}
	\end{subfigure}%\hfill
	\begin{subfigure}[t]{.33\linewidth}
        \begin{tikzpicture}
            \node [passive, text width={width("$\observation{t-1}$")}] (h0) at (0,0) {$\latent{t-1}$}; 
            \node [passive, text width={width("$\observation{t-1}$")}] (h1) at (2,0) {$\latent{t}$}; 
            \node [passive, text width={width("$\observation{t-1}$")}] (x0) at (0,-2) {$\observation{t-1}$};
            \node [passive, text width={width("$\observation{t-1}$")}] (x1) at (2,-2) {$\observation{t}$};
            \draw[->] (h0) -- (h1); % , color=red
            \draw[->] (h0) -- (x0); % , color=red
            \draw[->] (x0) -- (h1); % , dashed, color=red
            \draw[->] (h1) -- (x1);
        \end{tikzpicture}
		\caption{GUM}
	\end{subfigure}%\hfill
	\caption{Conditional dependencies 
	in HMC, RNN, and GUM.
	The dashed (resp. solid) lines stand for deterministic (resp. probabilistic) dependencies.}
	\label{fig:similarities}
	\end{center}
\end{figure*}

As a result, 
the GUM, HMC and RNN models 
only differ via
the distribution $\p(x_{0:t}|h_{0:t})$.
In an HMC, $h_s$ only depends on $h_{s-1}$ given the past 
$(h_{0:s-1},
x_{0:s-1})$,
so $\p(x_{0:t}|h_{0:t})=\prod_{s=0}^t p(x_s|h_s)$; 
on the other hand,
$h_t$ is stochastic so $\p(x_{0:t})$ is not available in closed form.
By contrast, 
in a D-GUM or an RNN,  
$h_s$ is deterministic given the past 
$(h_{0:s-1},
x_{0:s-1})$,
so $\p(x_{0:t})$ is available in closed form; 
but $h_s$ also depends on $h_{s-1}$ so
$\p(x_{0:t}|h_{0:t})$ is difficult to interpret.
The graphical representation of the three models is displayed in Fig.~\ref{fig:similarities}.

Now that we have cast the HMC and the RNN in a common framework, we can address the comparison of their expressivity from the GUM perspective. More precisely, our objective is to set some assumptions which enable us to discuss on the distribution of the observation $\p(x_{0:t})$ associated to the GUM, and next to discuss on the restrictions on this distribution induced by \eqref{eqcle-HMM} and \eqref{eqcle-RNN}.

\section{Structure of the mapping 
$\theta \rightarrow
p_{\theta}(\observation{0:t})$: the linear and Gaussian case}
\label{structured-mapping}

\subsection{Linear and Gaussian GUM}
\label{linear-gaussian-GUM}
We now address
%propose a comparative study of
the expressivity of the HMC and RNN models.
In order to compare the observations pdf $p(\observation{0:t})$ 
induced by the HMC and RNN models,
we set ourselves in the general framework of GUM,
in which 
$p(\observation{0:t})$
is a marginal of
\eqref{def-GUM}.
Of course,
$p(\observation{0:t})=\int_{\latent{0:t}}p(\latent{0:t},\observation{0:t})\dx\latent{0:t}$
cannot, in most cases, be computed in closed form.
In order to be able to provide a compared analysis of the expressivity of those models
(and thus to understand,
when we reduce to the particular cases of HMC
and of RNN,
the role of stochastic vs. deterministic transitions: 
see \eqref{eqcle-HMM}
and \eqref{eqcle-RNN}),
we consider the simplified linear and Gaussian framework,
i.e. a GUM model in which the elementary factors in 
\eqref{def-GUM}
read
\begin{align}
p(\latent{0})&=\gaussienne(\latent{0};0;\vi),
\label{eq:guminitiale}\\
p(\latent{t} |\latent{t-1},\observation{t-1}) 
&= \gaussienne(\latent{t};\thh\latent{t-1}+\txh\observation{t-1}; \vh),
\label{eq:gumtransition}\\   
p(\observation{t} |\latent{t})
&= \gaussienne(\observation{t}; \thx\latent{t}; \vx),
\label{eq:gumemission}
\end{align}
in which
$\latent{t}$ is an $n$-dimensional vector and
$\observation{t}$ is a scalar; so
$a$, $b$ and $c$ 
are respectively
$n \times n$,
$1 \times n$,
$n \times 1$,
$\eta$ and $\alpha$
are $n \times n$ covariance matrices,
and $\beta \geq 0$.
$\theta = 
(a, b, c, \alpha, \beta, \eta)$
is the parameter of the model.

It is easy to see that in model 
\eqref{def-GUM}
\eqref{eq:guminitiale}-\eqref{eq:gumemission},
the joint pdf
$p(\observation{0:t})$
is a zero-mean multivariate
Gaussian density,
which is fully characterized 
by its covariance matrix.
Let $\vi_t$ be the covariance matrix of $\latent{t}$ (we will later see how $\vi_t$ depends on $\vi_0$ and on time $t$). We have
$\var{\observation{t}} = \vx + \thx\vi_t\thx^T$,
and, for all $t\in\bN$, $k\in\bN^*$,
\begin{equation}
\cov{\observation{t}}{\observation{t+ k}} =
b
(\thh+\txh\thx)^{k-1}
(\underbrace{\thh\vi_t\thx^T+\txh(\vx + \thx\vi_t\thx^T)}_{N_t}).
\label{eq:RMC}
\end{equation}
Due to $\vi_t$ and factor $N_t$,
$\var{\observation{t}}$ and
$\cov{\observation{t}}{\observation{t+ k}}$ 
\textit{a priori} depend on time $t$.
In order to simplify the analysis
(see \S \ref{expressivity-GUM-HMM-RNN} below)
we first look for  simple sufficient conditions
yielding stationarity.

\subsection{Stationnarity}

First, it is easy to see that the matrix series $(\vi_t)_{t\in\bN}$ is defined by
\begin{equation}
% {\rm for \, all} t\in\bN^*,
\vi_{t+1} =
(\alpha +c\beta c^T)+(a+cb)\eta_t(a+cb)^T.
\label{eq:previrec}
\end{equation}
As a consequence, for all $t\in\bN$, 
$\vi_{t+1}-\vi_t =$
$(\thh+\txh\thx)^t\left[\vi_1-\vi_0\right](\thh+\txh\thx)^{tT}.
$
The series $(\vi_t)$ is thus constant if  
\begin{align}
    \vi_0=\vi_1=\vi.\label{eq:vicst}
\end{align}
Assumption \eqref{eq:vicst} implies in turn 
that 
$\var{\observation{t}}$ and $\cov{\observation{t}}{\observation{t+ k}}$ 
no longer depend on time,
so that 
$(\observation{t})_{t\in\bN}$ is a wide sense stationnary process.
Let us finally remark that under assumption \eqref{eq:vicst}, equation \eqref{eq:previrec} becomes:
\begin{align}
    \vi&=(\vh +\txh\vx \txh^T)  +  (\thh+\txh\thx) \; \vi \; (\thh+\txh\thx)^T.\label{eq:liapunov}
\end{align}
This equation in variable $\vi$ is meaningfull (recall that $\vi$ is a covariance matrix)
only if it admits a semi-definite positive ($\geq 0$) solution,
which implies
\cite{gevers1978innovations,brockett2015finite} 
that
\begin{align}
(\thh+\txh\thx)\text{ has all its eigenvalues in
}\{z\in\bC;|z|<1\}.
\label{eq:duo}
\end{align}

We will now assume that \eqref{eq:duo} and \eqref{eq:vicst} hold,
which implies that $\observation{t}$ is stationary.
Let us note that this stationarity assumption is reasonable, because under assumption \eqref{eq:duo},
the series $(\vi_t)$ converges when $t$ tends to infinity.
So $(\latent{t})_{t\in\bN}$ as well as  $(\observation{t})_{t\in\bN}$ are at least asymptotically stationary. 
\begin{remark}
\label{remark-particular-cases}
In the HMC case, we have the additional constraint $c=0$ since
$h_t$ does not depend on $x_{t-1}$ given $(h_{t-1},x_{t-1})$.
In the case of D-GUM, $\alpha=0$; if we also consider the full RNN case with its particular initial conditions (see \eqref{eqcle-RNN}), the constraint ${\rm Var}(h_1)=\eta=c{\rm Var}(x_0)c^T=c(\beta+b\eta b^T) c^T$ has also to be satisfied.
\end{remark}

\subsection{The mapping 
$\theta \rightarrow
p_{\theta}(\observation{0:t})$}
\label{mapping-cas-gaussien}%OLD: Image directe et image réciproque

Let $r_k=\cov{\observation{t}}{\observation{t+k}}$.
We observe that this covariance series has a very specific structure:
\begin{align}
\label{definition-r0}
&r_0 = \beta+b \, \vi \, b^T; \\
\nonumber
&{\rm for\,all\,} k\in\bN^*, \\
\label{definition-rk}
&r_k = 
\underbrace{\thx}_{H}
(\underbrace{\thh+\txh\thx}_{F})^{k-1}
(\underbrace{\thh\vi\thx^T+\txh(\vx + \thx\vi\thx^T)}_{N}).
\end{align}
Since this covariance series
$(r_k)_{k \in \bN}$ characterizes the distribution of
$p_{\theta}(\observation{0:t})$
for all $t$,
we now consider function
\begin{equation}
\label{direct-mapping-phi}
\phi :
\underbrace{(a,b,c,\alpha,\beta,\eta)}_{\theta}
\stackrel{
\eqref{definition-r0}-
\eqref{definition-rk}
}
{\longrightarrow}
\phi(\theta) =
(r_k)_{k \in \bN},
\end{equation}
in which $(r_k)_{k \in \bN}$
is given by 
\eqref{definition-r0}
\eqref{definition-rk}.
Since a study of the direct range of $\phi$
under the HMC or RNN constraints seems a difficult task,
we rather consider the inverse mapping.

Let us first observe that the factorized  structure of the covariance series
(i.e., there exists $(H,F,N)$
s.t.
$r_k = HF^{k-1}N$
for all $k\geq 1$)
is remarkable, 
and is directly related to system theory.
More precisely, the output of
any linear time invariant (LTI) state space system 
has a stationary factorized covariance series,
and conversely, any such series can be {\sl realized}
by an LTI system,
This second point is the topic of {\sl Stochastic realization theory (SRT)},
which indeed is of interest here 
since we shall look for parameters 
$\theta=(a,b,c,\alpha,\beta,\eta)$ s.t.
$\phi(\theta)=(r_k)_{k \in \bN}$
for a given 
$(r_k)_{k \in \bN}$.
Before we proceed to the analysis 
we thus briefly provide
a brief reminder of SRT
(the reader familiar with SRT can skip section \ref{app:trs}
and directly jump to section 
\ref{expressivity-GUM-HMM-RNN}).

\section{A short review of SRT}
\label{app:trs}

Let us briefly review some points from SRT
\cite{faurre1976stochastic, gevers1978innovations, faurre1979operateurs, gevers2006personal, caines2018linear}
which we will need in section
\ref{expressivity-GUM-HMM-RNN}.
SRT is a part of systems theory, which deals with  modeling, controlling and estimating dynamic systems (see e.g. \cite{chen1970introduction,kailath1980linear,chui2012signal}).
Before we proceed (see section \ref{annexe-srt})
we need to recall some algebraic facts from deterministic realization theory (DR).

\subsection{DRT}
\label{sec:A1}

Let us consider a linear discrete time system with state $\latent{t}$:
\begin{align}
\begin{cases}
\latent{t+1} &= F\latent{t} + Nu_t\\
\observation{t}   &= H\latent{t}
\end{cases}
,
\end{align}
where $F$ (resp. $N$, $H$) are 
$n\times n$ (resp. $n\times 1$, $1\times n$) matrices
(we only deal here with the case where observation $x_t$ and input $u_t$ are one-dimensional).
The mapping between input $u_t$ and output  $\observation{t}$ is given by the convolution equation
$\observation{t}=\sum_{k=1}^{+\infty}H_ku_{t-k}$,
where the lags $H_k$ of the impulse response 
(the so-called Markov parameters of the system)
satisfy
\begin{align}
    H_k=HF^{k-1}N\label{eq:A4}
\end{align}
for all $k\geq1$. Equivalently, the strictly causal transfer function
$H(z)=\sum_{k=1}^{+\infty}H_kz^{-k}$
can be written as
$H(z)=H(zI-F)^{-1}N$.

The DR problem consists in 
building three matrices $H,F,N$, 
with $F_{n\times n}$ of minimal  dimension, 
from the impulse response of the system,
i.e. move from the infinite representation $(H_k)_{k\in\bN^*}$
to the finite representation  $(H,F,N)$, with $F$ of minimal dimension.
The key tool for this problem is the infinite Hankel matrix
\begin{align}
\mathcal{H}_\infty=
\begin{bmatrix}
H_1 & H_2 & H_3 & \hdots \\
H_2 & H_3 &     & \\
H_3 &     &     & \\
\vdots &  &     & 
\end{bmatrix}.
\end{align}
From \eqref{eq:A4}, $\mathcal{H}_\infty$ factorizes as 
\begin{align}
\label{facto-Hinfty}
\mathcal{H}_\infty=
\begin{bmatrix}
H\\HF\\HF^2\\\vdots
\end{bmatrix}
.
[
N,FN,F^2N,...
],
\end{align}
and so has finite rank,
which moreover is equal to $n$
(the dimension of $F$)
if and only if (iff.) 
each factor is itself full rank $n$.
Conversely, if $\mathcal{H}_\infty$ has finite rank $n$, 
then it can be factorized 
as a product of two factors of dimensions
$(\infty \times n)$ and
$(n \times \infty)$,
both of them being of full rank $n$,
and due to the Hankel structure,
there exists $F_{n\times n},N_{n\times 1}, H_{1\times n}$ so that \eqref{facto-Hinfty}
(and thus \eqref{eq:A4})
is satisfied.
Moreover, from the proposition below, all minimal realizations of $H(z)$ are isomorphic:

\begin{proposition} 
\cite[proposition 3]{ho1966effective}
\label{prop:ho}
$(H_1,F_1,N_1)$ and $(H_2,F_2,N_2)$ are two minimal  realizations of $H(z)$
if and only if there exists $T$ invertible such that
$F_2=TF_1T^{-1}$,
$N_2=TN_1$ and
$H_2=H_1T^{-1}$.
\end{proposition}
Finally numerically efficient 
DR algorithms
have been proposed in \cite{de1975numerical,de1978numerical}.

\subsection{SRT}
\label{annexe-srt}

Let us now consider the state space system
\begin{align}
\begin{cases}
    \latent{t+1}&= F\latent{t}+u_t\\
    \observation{t}  &= H\latent{t}+v_t
\end{cases}\label{eq:A9},
\end{align}
where $\latent{0}$ is zero-mean and uncorrelated with $(u_t,v_t)$, and where $(u_t,v_t)$ is a zero-mean, uncorrelated, stationary random process with 
\begin{align}
\bE\left[
\begin{bmatrix}
    u_t\\v_t
\end{bmatrix}
.
\begin{bmatrix}
    u^T_{t'}&v^T_{t'}
\end{bmatrix}
\right]
&=
\begin{bmatrix}
Q&S\\S^T&R
\end{bmatrix}    
\delta_{t,t'}
\label{eq:A11}
\end{align}
and 
$\delta_{t,t'} = 1$
iff. $t=t'$.
Let us assume that $\{ x_t \}_{t \geq 0}$ 
is (wide sense) stationary
and purely non-deterministic.
This together with an observability condition on $(F,H)$
implies that 
$\{ h_t \}_{t \geq 0}$ is (wide-sense) stationary and purely non-deterministic as well.
Let  $P=\bE[\latent{t}\latent{t}^T]$; $P$ satisfies 
\begin{align}
    P=FPF^T+Q\label{eq:A12},
\end{align}
which in turn implies that $F$ has all its eigenvalues 
in the open unit disc.
Finally the covariance function of $\observation{t}$ is given by
\begin{align}
&r_0= \bE[\observation{t}^2] = R+HPH^T;
\label{eq:A121}\\
&    
\text{for all } k\!\in\!\bN^*,
r_k= 
\bE[\observation{t}\observation{t-k}]
\!=\! HF^{k-1}\!\underbrace{(FPH^T\!\!+S)}_{N}.
\label{eq:A13}
\end{align}
Starting from a covariance $(r_k)_{k\in\bN}$,
the SR problem consists in
building a minimal  "Markovian representation"
of $(\observation{t})_{t\in\bN}$,
i.e. a state-space system
\eqref{eq:A9}-\eqref{eq:A11}, with  $F$ of minimal dimension.

\subsubsection*{Step 1}

Thanks to the structure of function $(r_k)_{k\in\bN^*}$, we can as in section \ref{sec:A1} build a Hankel matrix
\begin{equation}
\hankel=
\begin{bmatrix}
\label{Rinfini}
r_1 \!\! & \!\! r_2 \!\! & \!\!r_3 \!\! & \!\!\hdots \\
r_2 \!\! & \!\! r_3 \!\! &      & \\
r_3 \!\! &     &      & \\
\vdots &  &      & 
\end{bmatrix}
=
\begin{bmatrix}
H\\HF\\HF^2\\\vdots
\end{bmatrix}
\begin{bmatrix}
N\!\!&\!\!FN\!\!&\!\!F^2N\!\!&\!\!\hdots
\end{bmatrix}
\end{equation}
which should be compared to factorization \eqref{facto-Hinfty}.
The first (and, in fact, "deterministic") step of a SR algorithm consists in building a minimal realization
$(H,F,N)$ of $(r_k)_{k\in\bN^*}$ (unique up to an  invertible matrix);

\subsubsection*{Step 2}

At this point, we dispose of $(H,F,N)$ but $N$ remains a function of $P$ and $S$ (see \eqref{eq:A13}),
and it remains to identify $Q$ and $R$.
This second step is more delicate for the problem must be solved under positivity constraints:
$P$ and
$\begin{bmatrix}
Q&S\\S^T&R
\end{bmatrix}
$
are covariance matrices and so must be semi-definite positive ($\geq 0$).
If these contraints were not statisfied, the solution would be meaningless.
Finally, the problem is as follows:
knowing $(H,F,N,r_0)$, we look for $(P,Q,R,S)$ such that 
\begin{align}
&\begin{bmatrix}
P&N\\N^T&r_0
\end{bmatrix}
-
\begin{bmatrix}
F\\H
\end{bmatrix}
P
\begin{bmatrix}
F^T&H^T
\end{bmatrix}
=
\begin{bmatrix}
    Q&S\\S^T&R
\end{bmatrix},\label{eq:A16}\\
% \end{align}
% \begin{align}
    &\;P>0,\label{eq:A17}\\
    &\begin{bmatrix}
    Q&S\\S^T&R    
    \end{bmatrix}
    \geq0,\label{eq:A18}
\end{align}
in which $>0$ stands for definite positive
(the constraint on $P$ should be, \textit{a priori}, that $P$ is \textit{semi-definite} positive,
but indeed it happens that any solution $P$ must be \textit{definite} positive \cite{faurre1979operateurs} (see theorem \ref{th:A7} below),
whence \eqref{eq:A17}).
Let us notice that equation \eqref{eq:A16} gives the covariance of
$(\latent{t+1},\observation{t})$.
Since $(u_t,v_t)$ is a white noise,
this covariance satisfies a (Ricatti) equation of the same kind as that satisfied by $P$ (equation \eqref{eq:A12}, 
which in fact is a submatrix of \eqref{eq:A16}).

System \eqref{eq:A16} can be seen as a system with three equations and four unknowns 
($P$, $Q$, $R$ and $S$),
or rather as a system with three equations and three unknowns 
($Q$, $R$ and $S$),
parameterized by $P$.
Finally, $P$ parameterizes 
solutions of the constrained system
\eqref{eq:A16}-\eqref{eq:A18}.
Let ${\cal P}$ be the set of parameters
\begin{equation}
\label{calP}
{\cal P} =
\{
P \,\, 
{\rm s. t.} \,\,
\eqref{eq:A16}-\eqref{eq:A18}
\,\, {\rm are} 
\,\, {\rm satisfied}
\}.
\end{equation}

\subsection*{Positive real lemma, positivity of 
$(r_k)_{k\in\bN}$,
structure of $\cP$}

A result known as the \textit{positive real lemma} 
(initially proved in the spectral domain)
connects the positivity of the series $(r_k)_{k\in\bN}$
(in other words, 
whether $(r_k)_{k\in\bN}$
is a 
{\sl covariance}
series)
to the existence of at least one solution to the constrained system
\eqref{eq:A16}-\eqref{eq:A18}.
Let us recall that the infinite series $(r_k)_{k\in\bN}$ is a covariance series iff. the Toeplitz form $\sum_{i,j=0}^mu_iu_jr_{|j-i|}$ is positive or null for all $m$,
i.e. iff. the associated Toeplitz matrix is semi-definite positive for all $m$.

\begin{lemma}[Positive real lemma \cite{faurre1979operateurs}]\label{lemma:lpr}
The series $(r_k)_{k\in\bN}$ is a covariance series iff. $\cP$ is non void.
\end{lemma}

We now consider the structure of $\cP$.

\begin{theorem}[\cite{faurre1979operateurs}]\label{th:A7}
The set $\cP$ is closed, convex, bounded and definite positive; it admits (for the usual order relation between symmetric matrices) a maximum $P^*$ and a minimum $P_*$.
\end{theorem}
Let us finally notice that there exist efficient algorithms for building elements of $\cP$ 
(see \cite{faurre1979operateurs,caines2018linear}).

\section{Expressivity of GUM, HMC and RNN}
\label{expressivity-GUM-HMM-RNN}

We are now ready to come back to mapping 
\eqref{direct-mapping-phi}
\eqref{definition-r0}
\eqref{definition-rk},
and first need to study the range of $\phi$.

\subsection{Algebraic properties induced by the factorizability and positivity constraints}
\label{sec:algebraic-induced}

The range of $\phi$ 
is strictly included into $\bR^{\bN}$,
since 
$(r_k)_{k \in \bN}$ in
\eqref{direct-mapping-phi}
is indeed a {\sl factorized} and
{\sl covariance} series.
As we now see,
it is possible to characterize this range, 
via algebraic tests which determine whether
a given real series
satisfies these two constraints.

\subsubsection{Factorizability}

First, factorization 
\eqref{definition-rk}
implies that the doubly infinite
Hankel matrix built on 
$(r_k)_{k \in \bN^*}$ 
factorizes as
\eqref{Rinfini}.
So the rank of $\hankel$ is finite 
and lower than or equal to $n$
(the dimension of $F$),
and is equal to $\dimh$ if and only if each factor has itself full rank $\dimh$.
In this case, 
$(H,F,N)$ is a so-called minimal (deterministic) realization of $(r_k)_{k \in \bN^{\star}}$
(see section
\ref{app:trs} for more details). 
One can show
(see \cite[proposition 3]{ho1966effective}
or section 
\ref{app:trs}) that all
minimal realizations 
are isomorphic:
$(H_1,F_1,N_1)$ and $(H_2,F_2,N_2)$ are two minimal realizations of $(r_k)_{k\in\bN^*}$ if and only if there exists $T_{1,2}$ invertible such that
\begin{align}
(H_2,F_2,N_2)=(H_1T_{12}^{-1},T_{12}F_1T_{12}^{-1},T_{12}N_1).
\label{eq:passage}
\end{align}

\subsubsection{Positivity} \label{positivity-positivereallemma}
Apart from being factorizable,
any sequence 
$(r_k)_{k \in \bN}
\stackrel
{\eqref{direct-mapping-phi}}
{=}
\phi(\theta)$
is also a covariance series,
which can be characterized either by 
the constraint that for all $k \in \bN$,
the Toeplitz matrix
with first row $[r_0,\cdots,r_k]$
is positive semi definite or,
equivalently,
that
$C(z) \stackrel{def}{=}
r_0 + 2
\sum_{k=1}^{\infty}
r_k z^k$
is a Carathéodory function, 
ie. has positive real part in the open unit disk
$\{z\in\bC;|z|<1\}$
(Carathéodory-Toeplitz theorem, see e.g.
\cite{akhiezer1965classical}).

In the context of this paper,
it is however more interesting to recall the positive real lemma, which relies on the factorizability constraint we just evoked. 
So assume that rank$(\hankel)$ is finite,
which enables to build a minimal set 
$(F,H,N)$ satisfying 
$r_k = H F^{k-1}N$ for all $k$, $k \geq 1$.
As we recalled in section \ref{app:trs},
positivity of the series
$(r_k)_{k \in \bN}$
is related to
whether there exists at least one
matrix $P >0$
satisfying 
\begin{equation}
\begin{bmatrix}
P&N\\N^T&r_0
\end{bmatrix}
-
\begin{bmatrix}
F\\H
\end{bmatrix}
P
\begin{bmatrix}
F^T&H^T
\end{bmatrix}
\geq 0,
\label{eq:C8}
\end{equation}
three unknowns ($Q,R$ and $S$) parameterized by $P$.
i.e. whether the set
${\cal P}$ defined in \eqref{calP}
is non void.

\subsection{
Compared expressivity of the three models}
\label{expressivity-gum-hmm-rnn}

Let us summarize section \ref{sec:algebraic-induced}.
\begin{itemize}
\item 
Starting from any real valued series $(r_k)_{k\in\bN}$, this series is factorizable
(i.e., there exists a triplet $(H,F,N)$ such that $r_k=HF^{k-1}N$ for all $k\in\bN^*$)
iff. the Hankel matrix $\hankel$ is finite rank;
the rank $\dimh$ of $\hankel$
is also the minimal dimension of any realization of $(r_k)_{k\in\bN}$;
\item
Starting from a factorizable series $(r_k)_{k\in\bN}$, this series is a covariance series if and only if there exists at least one matrix $P >0$
satisfying 
\eqref{eq:C8}.
\end{itemize}
This discussion is summarized in Fig. \ref{fig:C1} below, 
the South-West part of which 
is the range of function $\phi$ in 
\eqref{direct-mapping-phi}.

\begin{figure}[htbp]
\centering
    \includegraphics[scale=.2]{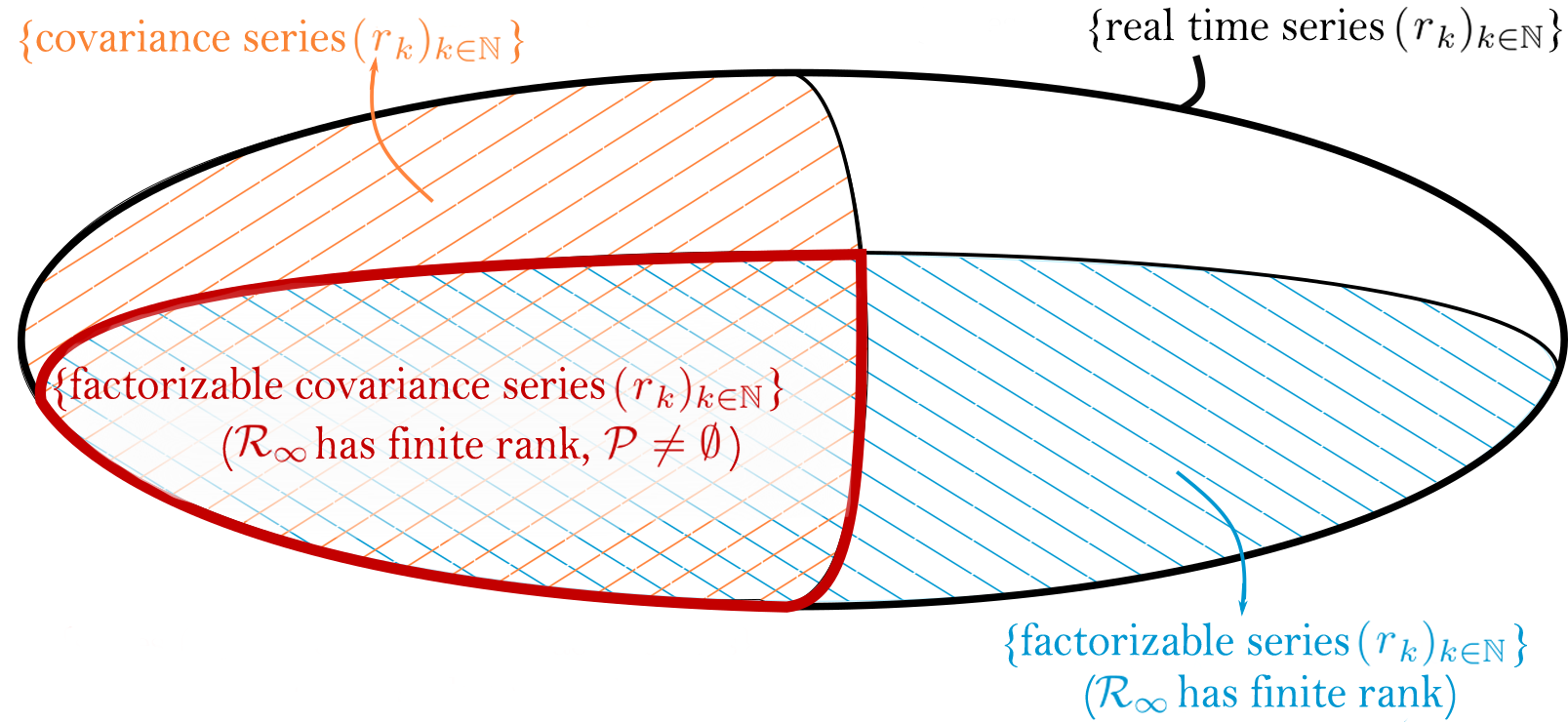}
    \caption{
This figure represents the set of all real times series. 
The series $(r_k)_{k\in\bN}$ which are factorizable  covariance series
is the South-West quarter of the figure (orange and blue lines).
Computing $\hankel$ enables to move from the full set to the Southern part, whereas the positive real lemma enables to move from the Southern part to the South-West quarter.
    }
    \label{fig:C1}
\end{figure}

We now study if any point of the South-West corner (i.e., any factorizable covariance function) can be realized by a GUM, 
an HMC and/or an RNN.

\subsubsection{Expressivity of GUM}

This question does not raise any particular difficulty.
Since $(r_k)_{k \in \bN}$ is a factorized covariance function,
it can be realized 
(see section \ref{app:trs})
by the state space system
\eqref{eq:A9}-\eqref{eq:A11}
for some  $(F,H,Q,R,S)$.
System 
\eqref{eq:A9}
can be rewritten 
(if $R \neq 0$)
as
\begin{eqnarray}
&&\begin{cases}
        \latent{t+1}&=a\latent{t}+c\observation{t} +u'_t\\
        \observation{t}&= b\latent{t}+v'-t
    \end{cases},
    \label{GUM-as-stsp1}
\\
&&
\bE\left[
    \begin{bmatrix}
        u'_t\\v'_t
    \end{bmatrix}
    .
    \begin{bmatrix}
        u_t^{'T}&v'_t
    \end{bmatrix}
    \right]
    =\begin{bmatrix}
         \alpha&0\\0&\beta
     \end{bmatrix}, 
     \label{GUM-as-stsp2}
\end{eqnarray}
in which 
\begin{align}
\label{transfo1}
a &= F-SR^{-1}H, b=H,
c=SR^{-1},
\\
\label{transfo2}
\begin{bmatrix}
\alpha&0\\0&\beta        
\end{bmatrix}
&=
\begin{bmatrix}
I&-SR^{-1}\\0&1
\end{bmatrix}
\underbrace{
\begin{bmatrix}
Q&S\\S^T&R
\end{bmatrix}
}
_
{\geq 0}
\begin{bmatrix}
I&0\\-R^{-1}S^T&1
\end{bmatrix};
\end{align}
Eq. \eqref{transfo2} 
ensures that 
$\begin{bmatrix}
\alpha&0\\0&\beta            
\end{bmatrix} \geq 0$ (and thus $\alpha \geq 0$).
Equations 
\eqref{GUM-as-stsp1}-\eqref{GUM-as-stsp2}
are a state space representation of 
\eqref{def-GUM}
\eqref{eq:guminitiale}-\eqref{eq:gumemission}.
In other words,
any point of the South-West corner can be realized by 
some linear and Gaussian GUM model.

\subsubsection{Expressivity of HMC}
We know that any factorizable covariance function 
$(r_k)_{k\in\bN}$ such that $\text{dim}(\hankel)=\dimh$
can be realized by a GUM of dimension $n$.
Starting from such a series, 
under which conditions does there exist an HMC
of the same degree $n$ which produces that same covariance series?
We have the following result (see Appendix \ref{proof-HMC} for a proof).
\begin{proposition}
\label{prop-hmc}
Let $(r_k)_{k\in\bN}$ a  factorizable covariance function and let $(H,F,N)$ a triplet
(with $F$ of minimal dimension $\dimh$)
produced by DR.
The series $(r_k)_{k\in\bN}$ can be realized by an HMC of dimension $n$ if and only if there exists $\tilde{P}$
(and thus $\tilde{Q}(\tilde{P})$ et $\tilde{R}(\tilde{P})$) such that
\begin{align}
&\begin{bmatrix}
\tilde{P}&N\\N^T&r_0
\end{bmatrix}
-
\begin{bmatrix}
F\\H
\end{bmatrix}
\tilde{P}
\begin{bmatrix}
F^T&H^T    
\end{bmatrix}
=
\begin{bmatrix}
    \tilde{Q}&0\\0&\tilde{R}
\end{bmatrix},
\label{eq:C21}\\
% \end{align}
% \begin{align}
    &\;\tilde{P}>0,\label{eq:C22}\\
    &\begin{bmatrix}
    \tilde{Q}&0\\0&\tilde{R}    
    \end{bmatrix}
    \geq0.\label{eq:C23}
\end{align}
\end{proposition}

\begin{remark}
Finally, let $\tilde{\cP}$ the set of solutions $\tilde{P}$ of the constrained problem  \eqref{eq:C21}-\eqref{eq:C23}.
One can note that $\tilde{\cP}$ is a  convex subset of $\cP$.
On the other hand, as compared to $\cP$, \eqref{eq:C21} yields the supplementary constraint \eqref{eq:C24}.
This equation can be satisfied only if $N\in\text{Span}(F)$.
Moreover, if $F$ is invertible, \eqref{eq:C24} also implies
\begin{align}
    H^TF^{-1}N>0.\label{eq:C25}
\end{align}
So if \eqref{eq:C24} and/or \eqref{eq:C25} is not satisfied, then the series $(r_k)_{k\in\bN}$
cannot be realized by an HMC of dimension $\dimh$.
\end{remark}

\subsubsection{Expressivity of D-GUM and RNN}

Similarly as the HMC, the study can be done from any triplet $(H,F,N)$ 
provided by the DR step. We have to take
into account the D-GUM constraint $\alpha=0$ (or $Q-SR^{-1}S^T=0$, see \eqref{transfo1}-\eqref{transfo2})
and the RNN constraint $\eta=c{\rm Var}(x_0)c^T$ (or 
$P=SR^{-1}r_0R^{-T}S^{T}$).  
\begin{proposition}
Let $(r_k)_{k\in\bN}$ a  factorizable covariance function and let $(H,F,N)$ a triplet (with $F$ of minimal dimension $\dimh$)
produced by DR. Let us note $\cP$
the set of solutions of system \eqref{eq:A16}-\eqref{eq:A18}.
Then $(r_k)_{k\in\bN}$ can be realized
by a D-GUM
if and only if there exists $\tilde{P}\in\cP$
such that 
\begin{equation}
\label{condition-D-GUM}
\tilde{P}-F\tilde{P}F^T- (N-F\tilde{P}H^T) (r_0-H\tilde{P}H^T)^{-1} (N-F\tilde{P}H^T)^T =  0 \text{.}
\end{equation}
If in addition $\tilde{P}$ satisfies 
\begin{equation}
\label{condition-RNN}
\tilde{P}= r_0(r_0-H\tilde{P}H^T)^{-2} (N-F\tilde{P}H^T) (N-F\tilde{P}H^T)^T \text{,}
\end{equation}
the covariance series can be produced by a
traditional RNN initialized to
$h_0=0$, with a linear activation function.
\end{proposition}
\begin{remark}
Note that by construction 
$\tilde{P}$
in 
\eqref{condition-RNN}
is a rank 1 $n \times n$ semi-definitive positive matrix,
and is positive definite only if $n=1$.
In other words,
a factorizable covariance series can be realized
by an RNN if the latent vector is monodimensional
and 
\eqref{condition-RNN} holds
(as we will check in the worked example below), but can never be realized by an RNN if $n>1$. 
\end{remark}

\subsection{A worked example (unidimensional case)}
We now illustrate the preceding section 
in the scalar case. We first look for conditions on 
the triplet $(H,F,N)$ such that a factorizable
series $(r_k)_{k\in\bN}$ is a covariance function.
We next give conditions  on this triplet to determine if the covariance series can be produced
by one of the generative models of this paper.

\subsubsection{SR step}
Let $(r_k)_{k\in\bN}$ be a covariance series, 
factorizable as $HF^{k-1}N$ for all $k\in\bN^*$ with  $H,F$ and $N$ scalar.
We assume that such a  triplet $(H,F,N)$ has been produced by DR of $(r_k)_{k\in\bN}$,
and we search for the scalar parameters $P,Q,R,S$ satisfying 
\eqref{eq:A16}-\eqref{eq:A18}.
Each of these equations becomes respectively
\begin{align}
    &\begin{cases}
         Q&=P(1-F^2)\\
         R&=r_0-PH^2\\
         S&=N-HFP
     \end{cases},\label{eq:665}\\
     &\;P\geq0,\label{eq:666}\\
     &\begin{cases}
          QR-S^2&\geq0\\
          Q&\geq0\\
          R&\geq0
      \end{cases}\label{eq:667}.
\end{align}
In particular,
\eqref{eq:667} corresponds to the semi-definite positive constraint \eqref{eq:A18}
when $Q,R$ and $S$ are scalar.

Let us build the set $\cP$ of positive numbers $P$ which satisfy this system. The second
inequality of \eqref{eq:667} is satisfied
when $F^2 \leq 1$;
by using \eqref{eq:665}, the first inequality of \eqref{eq:667} reads
\begin{align}
\label{eq:poly-2}
    \Xi_{/H^2}(P)\!\stackrel{\rm def.}{=}
    \!-H^2P^2\!\!+\!\![r_0(1\!-\!F^2)\!\!+\!\!2HFN]P\!\!-\!\!N^2 \!\geq\!0. %
\end{align}
Since polynomial $\Xi_{/H^2}$ is concave, 
one can show easily that \eqref{eq:poly-2}
admits a solution provided
\begin{equation}
\label{eq:cond_B}
\frac{r_0(F-1)}{2} \leq HN \leq \frac{r_0(F+1)}{2} \text{,}
\end{equation}
and that 
$\cP$ is included in $[P_{/H^2,1},P_{/H^2,2}]$ with
\begin{align}
\label{P1P2a}
P_{/H^2,i}&=\frac{(2HFN+r_0(1-F^2))+(-1)^i\sqrt{\delta}}{2H^2},\\
\label{P1P2b}
\delta &= (1\!-\!F^2)(r_0(1+F)\!-\!2HN)(r_0(1\!-\!F)+2HN).
\end{align}
Moreover, constraints \eqref{eq:666} and $R\geq0$ in \eqref{eq:667} imply that
$P$ belongs to $[0,\frac{r_0}{H^2}]$;
but it can be checked that $[P_{/H^2,1},P_{/H^2,2}]\subseteq[0,\frac{r_0}{H^2}]$ 
so constraints $P\geq0$ and $R\geq0$ do not yield further interval restrictions.
Finally, factorizable series
$(r_k)_{k\in\bN}$ is a covariance function
if  $F^2 \leq 1 $ and $HN$ satisfies \eqref{eq:cond_B}; $\cP$ then coincides with
\begin{align}\label{eq:674}
\cP=[P_{/H^2,1},P_{/H^2,2}]
\end{align}
and is non void. It can also be
produced by a GUM whose parameters are deduced from \eqref{transfo1}-\eqref{transfo2}.
\begin{remark}
Finally one can show easily that $P_{/H^2,1}>0$ if $N>0$
(the case $P_{/H^2,1}=0$ is possible only if $\Xi_{/H^2}(0)=-N^2\geq0$, and so $N=0$,
which corresponds to the degenerate case of a series $(r_k)_{k\in\bN}$ which is null everywhere except at $k=0$ where it is equal to $r_0$),
so $\cP$ is a \textit{definite} positive set, which is in concordance with Faurre's theory 
\cite[theorem 7]{faurre1979operateurs}.
\end{remark}

\subsubsection{HMC case} As a consequence of 
Proposition \ref{prop-hmc},  a factorizable covariance series $(r_k)_{k\in\bN}$ can be produced
by an HMC if there exists $\tilde{P}$ in $\cP$
such that $\tilde{P}=N(HF)^{-1}$; equivalently,
$\tilde{P}$ has to satisfy $\tilde{P}=N(HF)^{-1}$
and $\tilde{P} \in (0,r_0 H^{-2}]$. So condition \eqref{eq:cond_B} for $HN$ becomes 
\begin{align*}
\begin{cases}
0   < HN  \leq r_0 F \text{,}   \quad & \text{if } F \geq 0 \\
r_0 F  \leq HN  <  0 \text{,}  \quad & \text{if } F \leq 0
\end{cases} \text{.}
\end{align*}

\subsubsection{D-GUM and RNN cases} Remember that
for a D-GUM, the first inequality of \eqref{eq:667}
becomes an equality.
So polynomial
$\Xi_{/H^2}$
in
\eqref{eq:poly-2}
is equal to zero,
and system 
\eqref{eq:665}-\eqref{eq:667}
admits two solutions,
$P_{/H^2,1}$ and $P_{/H^2,2}$. 
Consequently, as the GUM, a D-GUM can produce any covariance series, but requires less parameters since $\alpha=0$.

Finally, the additional RNN constraint becomes $P= r_0 S^2R^{-2}$. In the same time $P$ has to satisfy $P=P_{/H^2,1}$ or $P=P_{/H^2,2}$. Using elementary calculus, 
these new systems have a solution if
$HN= r_0 F$, or 
$HN= r_0 F (2F^2-1)$. 

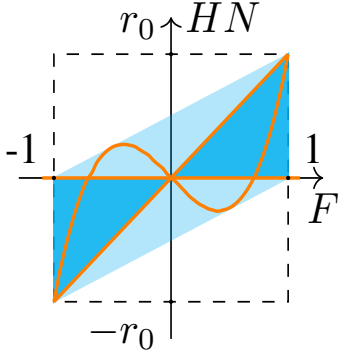
\begin{figure}[!ht]
\centering
\resizebox{5cm}{5cm}{
\begin{tikzpicture}
  \coordinate (O) at (0,0);
  \draw[->] (-1.3,0) -- (1.3,0) coordinate[label = {below:$F$}] (xmax);
  \draw[->] (0,-1.3) -- (0,1.3) coordinate[label = {right:$HN$}] (ymax);
  \draw[dashed] (-1,-1) -- (-1,1) -- (1,1) -- (1,-1) -- cycle ;
  \draw[cyan, fill=cyan, opacity=0.3] (-1,-1) -- (-1,0) -- (1,1) -- (1,0) -- cycle;
  \draw[cyan, fill=cyan, opacity=0.8] (-1,-1) -- (-1,0) -- (0,0) -- (1,1) -- (1,0) -- (0,0) -- cycle ;

  \draw[thick, orange] plot[smooth] coordinates {(-1.0,-1.0) (-0.96,-0.80) (-0.92,-0.63) (-0.88,-0.48) (-0.84,-0.34) (-0.8,-0.22) (-0.76,-0.11) (-0.72,-0.02) (-0.67,0.051) (-0.64,0.115) (-0.6,0.168) (-0.56,0.208) (-0.52,0.238) (-0.48,0.258) (-0.43,0.269) (-0.4,0.271) (-0.36,0.266) (-0.31,0.254) (-0.28,0.236) (-0.24,0.212) (-0.19,0.183) (-0.16,0.151) (-0.12,0.116) (-0.07,0.078) (-0.04,0.039) (0.0,-0.0) (0.040,-0.03) (0.080,-0.07) (0.120,-0.11) (0.159,-0.15) (0.199,-0.18) (0.24,-0.21) (0.28,-0.23) (0.320,-0.25) (0.360,-0.26) (0.400,-0.27) (0.439,-0.26) (0.48,-0.25) (0.52,-0.23) (0.56,-0.20) (0.600,-0.16) (0.640,-0.11) (0.679,-0.05) (0.72,0.026) (0.76,0.117) (0.8,0.224) (0.840,0.345) (0.880,0.482) (0.919,0.637) (0.96,0.809) (1.0,1.0)};
  \draw[thick, orange] (-1.1,0) -- (1.1,0) ;
  \draw[thick, orange] (-1,-1) -- (1,1) ;
  
  \filldraw (-1,0) circle (.3pt) node[anchor=south east] {-1};
  \filldraw (1,0)  circle (.3pt) node[anchor=south west] {1};
  \filldraw (0,-1) circle (.3pt) node[anchor=north east] {$-r_0$};
  \filldraw (0,1)  circle (.3pt) node[anchor=south east] {$r_0$};
\end{tikzpicture}}
\caption{Expressivity of RNN, HMC, D-GUM and GUM 
with regards to parameters $F$ and $HN$ in the scalar case. The parallelogram (blue and cyan) coincides with the factorizable covariance series $r_k=F^{k-1}HN$. Such series can be produced by a GUM or a D-GUM. The blue (resp. orange) area (resp. curves) coincides with the value of $F$ and $HN$ which can be taken by the HMC (resp. the RNN).} 
\label{fig:cartography}
\end{figure}

A graphical representation of the expressivity
of each generative model in function of parameters 
$F$ and $HN$ is given in Fig. \ref{fig:cartography} below
(these 1-dimensional results coincide with those obtained in 
\cite{salaun2019comparing}
by using the Caratheodory theorem).

\section{Conclusion}
In this paper we adressed a comparative study of HMC and RNN, 
which are familiar tools for predicting time series.
Even though both tools were developed in different communities, 
we first showed that they indeed share close features when the RNN is turned into a generative model,
and thus when HMC and RNN are considered as two latent variables probabilistic models with close enough (conditional) independence structures.
Under this framework, both structures can be seen as two different particular instances of a common generative unified model.
We next compared both models from the point of view of {\sl expressivity}, i.e. the relative complexity of the joint probability distribution of an observations sequence, 
induced by the underlying latent variables.
By contrast with previous studies, which were of an experimental nature,
our approach consisted in thoroughly quantifying the modeling power of both models.
To that end we considered the linear and Gaussian assumption, 
which induces that
the probability distributions of an observations sequence produced by each model are characterized by structured covariance series, 
which enabled us to call for SRT.
Finally we provided implicit conditions under which a given covariance series (and thus a given probability distribution)
can be realized by a GUM, an HMC and/or a D-GUM or an RNN. These implicit conditions turn to an explicit cartography of the models in the mono-dimensional case.

\appendix

\subsection{Inference algorithms in HMC}
\label{Computations-HMM}
\subsubsection{Computing the likelihood}
As we recalled in section \ref{sec:markov},
being able to compute and maximize the likelihood is a key factor for choosing a probabilistic model.
The likelihood can be computed from the predictive likelihoods
$p_\parameters(\observation{s}|\observation{0:s-1})$.
In model  \eqref{eq:joint-hmc},
for all $s$, $0 \leq s \leq t$, we have
\begin{align}
&p_\parameters(\observation{s}|\observation{0:s-1})
= \nonumber \\ &
\int
\underbrace{
p_\parameters(\latent{s-1}|\observation{0:s-1})
}_{\text{filtering pdf}}
\underbrace{
p_\parameters(\latent{s}|\latent{s-1})
p_\parameters(\observation{s}|\latent{s})
}_{\text{HMC transition pdf}}
\dx\latent{s-1:s},\label{eq:310}
\end{align}
where 
$p_\parameters(\observation{s}|\latent{s})$ and 
$p_\parameters(\latent{s}|\latent{s-1})$
are the elementary factors 
in 
\eqref{eq:joint-hmc}.
On the other hand, the filtering pdf
$p_\parameters(\latent{s-1}|\observation{0:s-1})$ 
can be computed recursively:
\begin{align}
p_\parameters(\latent{s}|\observation{0:s}) %&=
\ &=  \frac{p_\parameters(\observation{s}|\latent{s})}{p_\parameters(\observation{s}|\observation{0:s-1})} \times \nonumber \\ &
\int
p_\parameters(\latent{s}|\latent{s-1})p_\parameters(\latent{s-1}|\observation{0:s-1})
\dx\latent{s-1}.\label{eq:311}
\end{align}
So equations 
\eqref{eq:310} and \eqref{eq:311} 
enable to compute the predictive pdf
$p_\parameters(\observation{s}|\observation{0:s-1})$ 
and the filtering pdf  $p_\parameters(\latent{s}|\observation{0:s})$
recursively.
Given the initial pdf $p_\parameters(\latent{0})$, 
we first compute 
$p_\parameters(\observation{s}|\observation{0:s-1})$ 
from 
$p_\parameters(\latent{s-1}|\observation{0:s-1})$ 
via \eqref{eq:310}.
We next compute $p_\parameters(\latent{s}|\observation{0:s})$
from 
$p_\parameters(\latent{s-1}|\observation{0:s-1})$, 
$p_\parameters(\observation{s}|\observation{0:s-1})$
and the HMC transition pdfs.
Note that it is the HMC structure 
\eqref{eq:joint-hmc}
that enables
this likelihood calculation,
at least theoretically
(in practice, the integrals in equations \eqref{eq:310} and \eqref{eq:311} 
can be difficult to compute,
see section \ref{HMC-pratique} 
for further discussion).

\subsubsection{Learning}
In latent variables models
(as is the case here)
computing the maximum likelihood estimate is difficult,
and one generally resorts to approximations.
In particular, 
the EM algorithm 
is an iterative learning method 
which runs as follows.
At step $i$,
we first compute,
under parameter $\parameters_i$,
the expected log-likelihood
given observations 
$\observation{0:t}$:
\begin{align}
& \bE_{\parameters_i}\left[\log p_\parameters(\observation{0:t},\latent{0:t})|\observation{0:t}\right] \nonumber \\ &
=
\int%_{\latent{0:t}}
p_{\parameters_i}(\latent{0:t}|\observation{0:t}) \log p_\parameters(\observation{0:t},\latent{0:t})\dx\latent{0:t}\label{eq:E}
\end{align}
\begin{equation}
=
\sum_{s=1}^t
\int%_{\latent{s-1},\latent{s}} p_{\parameters_i}(\latent{s-1},\latent{s}|\observation{0:t})
\left[ \log p_\parameters(\observation{s}|\latent{s})
                    + \log p_\parameters(\latent{s}|\latent{s-1})\right] p_\parameters (h_{s-1:s}|x_{0:t})
\dx\latent{s-1:s}\nonumber
\end{equation}
\begin{equation}
+
\int%_{\latent{0}} p_{\parameters_i}(\latent{0}|\observation{0:t})
\log\left(p_\parameters(\observation{0}|\latent{0})p_\parameters(\latent{0})\right)
\dx\latent{0},\label{eq:E2}
\end{equation}
\noindent
next we update this parameter
$\parameters_{i}$
$\rightarrow$
$\parameters_{i+1}$ 
by maximizing 
\begin{align}
\parameters_{i+1} = \underset{\parameters}{\text{argmax }}
\bE_{\parameters_i}\left[\log p_\parameters(\latent{0:t},\observation{0:t})|\observation{0:t}\right].
\label{eq:M}
\end{align}
Equations \eqref{eq:E} and \eqref{eq:M} are respectively the E and M steps of the EM algorithm.
In particular,
the E step is feasible if factors
$p_{\parameters_i}(\latent{s-1},\latent{s}|\observation{0:t})$
and
$p_{\parameters_i}(\latent{0}|\observation{0:t})$
can be computed.
As is well known, the algorithm
ensures that the likelihood increases with the iterations:
for all $i$,
$
p_{\parameters_i}(\observation{0:t})\leq p_{\parameters_{i+1}}(\observation{0:t}).
$
Stronger theoretical guarantees are available under further conditions
\cite{wu1983convergence}\cite{balakrishnan2017statistical}.
%there is no guarantee of convergence towards a global optimum.

\subsubsection{Practical considerations}
\label{HMC-pratique}

In practice, computing the likelihood 
and maximizing it via the EM algorithm
depend on the model assumptions.
We can distinguish three different cases.

\begin{flushleft}
{\sl Case 1: Linear and Gaussian state space systems.}
\end{flushleft}

Assume that %the series 
$\observation{0:t}$ 
and $\latent{0:t}$
take continuous values,
and that the transition pdfs
$p_\parameters(\latent{s+1}|\latent{s})$ and $p_\parameters(\observation{s}|\latent{s})$
are linear and Gaussian:
$
\latent{s+1} =
\Fs\latent{s} + \us$,
$\observation{s} =
\Hs\latent{s} + \vs$
where $\Fs$ and $\Hs$ are matrices and $\latent{0}$ and
$(\us,\vs)$ 
are Gaussian independent random vectors.
Under such assumptions
all pdfs of interest are indeed Gaussian,
so propagating them through time 
reduces to propagating their parameters.

Computing the likelihood in this model
can be done via 
an iterative algorithm
known as the Kalman filter (KF),
introduced in the control community in the 1960's  \cite{kalman1960new, kalman1961new, ho1964bayesian}
and heavily studied since then \cite{anderson2012optimal,kailath2000linear, meinhold1983understanding}.
The KF enables 
to compute efficiently the filtering pdf 
$p_\parameters(\latent{s}|\observation{0:s})$ for any $s$.
Similarly, 
one can show that the parameters of
$p_\parameters(\latent{s-1},\latent{s}|\observation{0:t})$ 
(see \eqref{eq:E2}) 
and of the smoothing pdf  $p_\parameters(\latent{s}|\observation{0:t})$ (which are also Gaussian)
can be computed via backward propagation 
\cite{shumway1982approach},
which enables an efficient implementation of the EM algorithm.

\begin{flushleft}
{\sl Case 2: continuous states (general case).}
\end{flushleft}

In the general case 
(non linear transition pdfs and/or non Gaussian noise),
exact computing is not available and one needs to resort to approximations.
Approximation methods include 
the extended KF,
i.e. a KF in a linearized model
\cite{chen1993approximate} \cite{anderson2012optimal},
and the unscented KF,
which propagates an approximation of 
the one- and second-order moments of the pdfs of interest
\cite{julier2004unscented}
\cite{julier2000new}
\cite{menegaz2015systematization}.

Particle filtering
(or sequential Monte Carlo) methods
are another class of approximate solutions
\cite{doucet2001sequential}
\cite{arulampalam2002tutorial}
\cite{chopin-papaspiliopoulos-SMC},
which consist in propagating a random, discrete approximation of $p_\parameters(\latent{0:t}|\observation{0:t})$,
via an importance sampling mecanism
with resampling
\cite{kahn1953methods}.
Let us start from
$\hat{p}_\parameters(\latent{0:t}|\observation{0:t}) =$
$
\sum_{i=1}^N w_{t}^{(i)}\delta_{\latent{0:t}^{(i)}}(\latent{0:t}),
$
where $\delta$ is the Dirac mass
and $w_{t}^{(i)}$ are a normalized set of weights.
The weighted trajectories
$\{\latent{0:t}^{(i)}, w_{t}^{(i)}\}_{i=1}^{N}$
are propagated via three steps.
For all $s\in\bN$, 
the $i^{\rm th}$ particle is sampled from
a conditional importance pdf $q$:
\begin{align}
\tilde{h}^{(i)}_{s+1}\sim q(\latent{s+1}|\latent{s}^{(i)}).\label{eq:ispropag}
\end{align}
Next we compute its unnormalized weight
\begin{align}
\tilde{w}_{s+1}^{u,(i)} =  w_{s}^{(i)} p_\parameters\left(\tilde{h}_{s+1}^{(i)}|\latent{s}^{(i)}\right)
p_\parameters\left(\observation{s+1}|\tilde{h}_{s+1}^{(i)}\right)
/
q\left(\tilde{h}_{s+1}^{(i)}|\latent{s}^{(i)}\right)
,
\label{eq:isweight}
\end{align}
which is normalized as
$
\tilde{w}_{s+1}^{(i)} = \tilde{w}_{s+1}^{u,(i)}
/
\sum_{j=1}^N \tilde{w}_{s+1}^{u,(j)}.
$
Finally the trajectories can be resampled, i.e.
$
h_{0:s+1}^{(i)}\sim 
$
$
\sum_{j=1}^N \tilde{w}_{s+1}^{(j)}\delta_{(\latent{0:s}^{(j)},\tilde{h}_{s+1}^{(j)})}(\latent{0:s+1}),
$
and given new weights 
$
w_{s+1}^{(i)} = \frac{1}{N}.
$
This optional resampling step
keeps a larger proportion of trajectories with strong weights,
to the detriment of those of low weight. 
If the trajectories are not resampled, then
$\tilde{h}_{s+1}^{(i)}$ 
(resp. $\tilde{w}_{s+1}^{(i)}$)
reduces to  $\latent{s+1}^{(i)}$ 
(resp. $w_{s+1}^{(i)}$).
With or without resampling, 
the procedure is repeated from \eqref{eq:ispropag}.

The unnormalized weights computed in \eqref{eq:isweight} enable in turn 
to compute an approximation of the predictive likelihood:
$
\hat{p}_\parameters(\observation{t}|\observation{0:t-1}) = \frac{1}{N} \sum_{i=1}^N \tilde{w}_{t}^{u,(i)}
$,
from which an estimate of the likelihood is computed from \eqref{eq:facto}.
Finally 
$\hat{p}_\parameters(\latent{0:t}|\observation{0:t})$ also provides an approximation of \eqref{eq:E}:
\begin{displaymath}
\hat{\bE}_{\parameters'}\left[\log p_\parameters(\latent{0:t},\observation{0:t})|\observation{0:t}\right] 
= \sum_{i=1}^N w_{\parameters',t}^{(i)} \log \hat{p}_\parameters(\latent{0:t}^{(i)},\observation{0:t}),
\end{displaymath}
which still remains to be maximized
(notation $w_{\parameters',t}^{(i)}$ 
recalls that weights are built from parameter $\parameters'$, see \eqref{eq:isweight}).
In practice approximation $\hat{p}$ can be poor, in particular when $N$ is very small w.r.t. $t$.
As a possible rescue one can use
particle smoothing algorithms
\cite{kantas2015particle}
\cite{briers2010smoothing}
\cite{carvalho2010particle}
\cite{fearnhead2010sequential},
which aim at improving  the approximation of $p(\latent{s-1},\latent{s}|\observation{0:t})$
in \eqref{eq:E2}.

\begin{flushleft}
{\sl Case 3 : discrete latent states.}
\end{flushleft}

HMC with discrete latent states were introduced in the 1960's
\cite{baum1966statistical}
\cite{baum1967inequality}
\cite{forney1973viterbi}
and have been used in such fields as 
langage processing \cite{rabiner1986introduction}
\cite{rabiner1989tutorial}, 
bioinformatics \cite{koski2001hidden}
or digital communications 
\cite{viterbi1967error}
\cite{forney1973viterbi}.

In the discrete case, 
the problem is that computing the likelihood as
$
p_\parameters(\observation{0:t})=\sum_{\latent{0:t}} p_\parameters(\latent{0:t},\observation{0:t})
 \label{eq:marginalisation}
$,
i.e. via brute force marginalization 
of the full joint pdf,
is unfeasible due to the exponential cost.
The success of HMC comes from the fact that the computation of the likelihood $p_\parameters(\observation{0:t})$
can be performed in linear time.
Indeed the likelihood can be seen as another marginalized pdf:
\begin{align}
    p_\parameters(\observation{0:t}) = \sum_{\latent{t}}\underbrace{p_{\parameters}(\latent{s},\observation{0:s})}_{\alpha(\latent{s})},
\end{align}
in which pdfs $\alpha(\latent{s})$
can be computed recursively in linear cost in the {\sl forward} time direction:
\begin{align}
\alpha(\latent{0})   &= p_\parameters(\latent{0})p_\parameters(\observation{0}|\latent{0})\\
\alpha(\latent{s+1}) &= p_{\parameters}(\observation{s+1}|\latent{s+1})\sum_{\latent{s}}
p_{\parameters}(\latent{s+1}|\latent{s})\alpha(\latent{s}).\label{eq:235}
\end{align}
Note that $\alpha(\latent{s})$ is proportional to the filtering probability mass function,
and that \eqref{eq:235} is the discrete analog of \eqref{eq:311}.
As for the predictive likelihood, it reads
\begin{align}
\nonumber
p(\observation{s+1}|\observation{0:s}) &= \frac{\sum_{\latent{s+1}}p_\parameters(\observation{s+1}|\latent{s+1})
\sum_{\latent{s}}p(\latent{s+1}|\latent{s})\alpha(\latent{s})
}{\sum_{\latent{s}} \alpha(\latent{s})}.
\end{align}

From \eqref{eq:E2} (where integrals become sums), 
running the EM algorithm requires calculating,
for all $s$, $0 \leq s \leq t$,
pdf 
$p_\parameters(\latent{s-1},\latent{s}|\observation{0:t})$, 
which is proportional to 
\begin{align}
p_\parameters(\latent{s-1},\latent{s},\observation{0:t}) = 
\underbrace{p_\parameters(\observation{s+1:t}|\latent{s})}_{\beta(\latent{s})}
\alpha(\latent{s-1})
p_\parameters(\latent{s}|\latent{s-1})
p_\parameters(\observation{s}|\latent{s}).\label{eq:250}
\end{align}
In particular, pdf $p_\parameters(\latent{s-1},\latent{s},\observation{0:t})$ 
depends on the backward pdfs
$\beta(\latent{s})  = p_{\parameters}(\observation{s+1:t}|\latent{s})
$
which, similarly to pdfs $\alpha(\latent{s})$,
can be computed recursively at linear cost, 
but in the reverse time direction (whence the term {\sl backward}):
\begin{align}
\beta(\latent{t}) &= 1 \\    
\beta(\latent{s}) &= \sum_{\latent{s+1}}\beta(\latent{s+1})p_{\parameters_k}(\observation{s+1}|\latent{s+1})
p_{\parameters}(\latent{s+1}|\latent{s}).
\end{align}
The recursive calculation of fonctions $\alpha(\latent{s})$ and $\beta(\latent{s})$, 
for all $s$, 
is the so called forward-backward algorithm
\cite{baum1966statistical}
\cite{baum1967inequality}
\cite{rabiner1989tutorial}.
Finally from \eqref{eq:250} we have
\begin{align}
p_\parameters(\latent{s-1},\latent{s}|\observation{0:t}) 
&= \frac{
\beta(\latent{s})
\alpha(\latent{s-1})
p_\parameters(\latent{s}|\latent{s-1})
p_\parameters(\observation{s}|\latent{s})
}
{\sum_{\latent{s-1},\latent{s}} 
\beta(\latent{s})
\alpha(\latent{s-1})
p_\parameters(\latent{s}|\latent{s-1})
p_\parameters(\observation{s}|\latent{s})
}.\label{eq:255}
\end{align}
It remains to maximize w.r.t.
$p_\parameters(\latent{t}|\latent{t-1})$
and  
$p_\parameters(\observation{t}|\latent{t})$.
Computing  $p_\parameters(\latent{s-1},\latent{s},\observation{0:t}) $
enables to update these pmfs / pdfs 
in the M step of the EM algorithm;
this version of the EM algorithm,
applied to discrete latent states HMC, 
is called the Baum-Welch algorithm 
\cite{baum1970maximization, rabiner1989tutorial}.
Finally observe that the forward-backward
algorithm enables to compute the smoothing pmf
(for a given, fixed parameter $\parameters$), since from  \eqref{eq:255} we get
$p_{\parameters}(\latent{s}|\observation{0:t})=\frac{\alpha(\latent{s})\beta(\latent{s})}{\sum_{\latent{s}}
\alpha(\latent{s})\beta(\latent{s})}.
$

\subsection{Proof of Proposition \ref{prop-hmc}}
\label{proof-HMC}
According to Remark \ref{remark-particular-cases},
the problem reduces to studying whether among the set of all solutions, 
there exists at least one such that $c=0$, 
and thus (see \eqref{transfo1}) $S=0$.
So we need to solve \eqref{eq:C8}
with the additional constraint
\begin{align}
    \label{eq:C24}
    N=FPH^T,
\end{align}
whence \eqref{eq:C21}. However, this raises the following question: although the set $\cP_1$ in which we look for an HMC solution is built from \textit{a} triplet $(H_1,F_1,N_1)$, produced by the DR step, this triplet is not unique;  if $\cP_1$ had no HMC solution, could another set $\cP_2$, built from another triplet $(H_2,F_2,N_2)$, nevertheless contain an HMC solution?

We thus need to study the relation between $\cP_1$ and $\cP_2$.
Since $(H_1,F_1,N_1)$ is of minimal degree,
from \eqref{eq:passage}
any other minimal degree solution $(H_2,F_2,N_2)$ 
can be computed from $(H_1,F_1,N_1)$ 
via an invertible matrix $T_{12}$.
Let $P_1$ be an element of $\cP_1$.
By pre- (respectively post-) multiplying equation \eqref{eq:C8}, with parameters $(H_1,F_1,N_1)$,
by $\begin{bmatrix}
T_{12}&0\\0&1         
\end{bmatrix}$
(respectively 
$\begin{bmatrix}
T_{12}&0\\0&1         
\end{bmatrix}^T$),
one can show easily that
\begin{equation}
\label{congruence-P1-P2}
\cP_2 
= \{T_{12}P_1T_{12}^T\text{ ; }P_1\in\cP_1\},
\end{equation}
which we denote simply by the set equation 
$\cP_2=T_{12}\cP_1T_{12}^T$.
This is summarized by 
Figure \ref{fig:C2}
below.

\begin{figure}[htbp!]
\begin{center}
    \includegraphics[scale=.3]{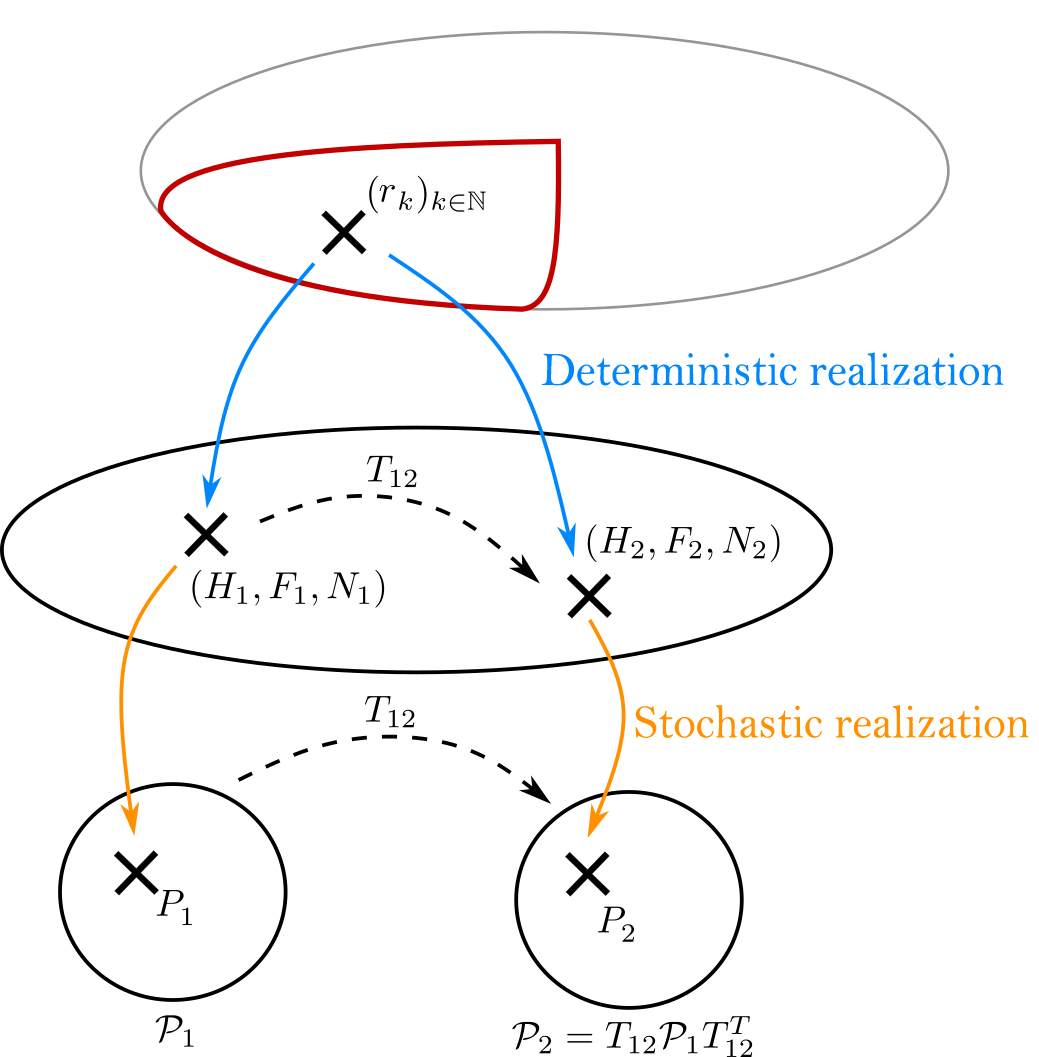}
\end{center}
\caption{
Starting from a factorizable covariance series $(r_k)_{k\in\bN}$
(see figure \ref{fig:C1}), the  deterministic realization step (in blue) consists in finding a  triplet $(H,F,N)$ representing the series under study. This step provides one solution 
out of an infinity of solutions to this problem.
These solutions are isomorphic and it suffices to know the appropriate invertible matrix $T_{12}$ to move from a given solution $(H_1,F_1,N_1)$ to another $(H_2,F_2,N_2)$ (dashed arrow).\\
The SR step (in orange) amounts to finding a state-space system modeling function 
$(r_k)_{k\in\bN}$ from the  triplet $(H,F,N)$ obtained at the previous step.
A triplet $(H_i,F_i,N_i)$ leads to a set of solutions $\cP_i$. 
These sets are also isomorphic and $T_{12}$ suffices for moving from $\cP_1$ to $\cP_2$, respectively
obtained from $(H_1,F_1,N_1)$ and $(H_2,F_2,N_2)$ (dashed arrow).
}
    \label{fig:C2}
\end{figure}

Let now a triplet $(H_1,F_1,N_1)$ and a solution $P_1\in\cP_1$, such that
\begin{align}
    N_1=F_1P_1H_1^T.\label{eq:NFPH}
\end{align}
This matrix $P_1$ is thus an HMC solution of $(r_k)_{k\in\bN}$.
Equation \eqref{eq:NFPH} is equivalent to
\begin{align}
     \underbrace{T_{12}N_1}_{N_2}=
     \underbrace{{T_{12}}F_1T_{12}^{-1}}_{F_2}
     \underbrace{T_{12}P_1T_{12}^T}_{P_2\in\cP_2}
     \underbrace{T_{12}^{-T}H_1^T}_{H_2^T};
\end{align}
so
from \eqref{eq:passage} 
and \eqref{congruence-P1-P2},
we see that
$P_2=T_{12}P_1T_{12}^T$ is one HMC element belonging to set $\cP_2$. 
In other words, if there exists an HMC element in the set $\cP_1$ associated to a triplet $(H_1,F_1,N_1)$ from the equivalence class produced by the DR step, then any set $\cP_2=T_{12}\cP_1T_{12}^T$ (with $T_{12}$ an arbitrary invertible matrix) also contains an HMC solution.
Similarly, if there is no such element in $\cP_1$, then no set $\cP_2=T_{12}\cP_1T_{12}^T$ will contain a solution either.
Finally it suffices to look for an HMC solution in the set $\cP$ produced by the SR algorithm, without bothering any longer of the other elements in the  equivalence class.
\bibliographystyle{ieeetr}
\bibliography{manuscrit}
\end{document}